\documentclass{statsoc}
\usepackage{amssymb}
\usepackage{amsmath}
\usepackage{graphicx}
\usepackage{natbib}
\usepackage{url}
\usepackage{setspace}

\newcommand{\e}{{\rm e}}
\newcommand{\dx}{{\rm d}x}
\newcommand{\dt}{{\rm d}t}

\newcommand{\onehalfspaceb}{\def\baselinestretch{1.5}\large\normalsize}

\title[On the Frequency of Severe Terrorist Events]{On the Frequency of Severe Terrorist Events}
\author[Clauset, Young and Gleditsch]{Aaron Clauset\footnotetext{The journal version of this pre-print appeared as ``On the Frequency of Severe Terrorist Events,'' {\em Journal of Conflict Resolution}, {\bf 51}(1): 58 -- 88 (2007), which can be found at {\tt http://jcr.sagepub.com/cgi/content/abstract/51/1/58}. \vspace{2mm}
    \\
    {\em Address for correspondence:}
    Aaron Clauset, 1399 Hyde Park Rd., Santa Fe NM, 87501 USA. \\
    E-mail: {\tt aaronc@santafe.edu}, {\tt young@cs.unm.edu}, {\tt ksg@essex.ac.uk}}}

\address{Santa Fe Institute, Santa Fe, NM, USA and \\
         University of New Mexico, Albuquerque, NM, USA.}
\author[Clauset, Young and Gleditsch]{Maxwell Young}
\address{University of New Mexico, Albuquerque, NM, USA.}
\author[Clauset, Young and Gleditsch]{Kristian Skrede Gleditsch}
\address{University of Essex, Wivenhoe Park, Colchester, UK and \\
       Centre for the Study of Civil War, Oslo, Norway.}


\begin{document}

\begin{abstract}
In the spirit of Richardson's original~\citeyearpar{Richardson48} study of the statistics of deadly conflicts, we study the frequency and severity of terrorist attacks worldwide since 1968. We show that these events are uniformly characterized by the phenomenon of {\em scale invariance}, i.e., the frequency scales as an inverse power of the severity, $P(x)\propto x^{-\alpha}$. We find that this property is a robust feature of terrorism, persisting when we control for economic development of the target country, the type of weapon used, and even for short time-scales. Further, we show that the center of the distribution oscillates slightly with a period of roughly $\tau\approx13$ years, that there exist significant temporal correlations in the frequency of severe events, and that current models of event incidence cannot account for these variations or the scale invariance property of global terrorism. Finally, we describe a simple toy model for the generation of these statistics, and briefly discuss its implications.
\end{abstract}

\onehalfspaceb
\fontsize{11}{11}
\selectfont

\section{Introduction}
Richardson first introduced the concept of {\em scale invariance}, i.e., a power-law scaling between dependent and independent variables, to the study of conflict by examining the frequency of large and small conflicts, as a function of their severity~\citep{Richardson48}. His work demonstrated that for both wars and small-scale homicides, the frequency of an event scales as an inverse power of the event's severity (in this case, the number of casualties). Richardson, and subsequent researchers such as~\cite{Cederman03}, have found that the frequency of wars of a size $x$ scales as $P(x)\propto x^{-\alpha}$, where $\alpha\approx 2$ and is called the scaling exponent. Recently, similar power-law statistics have been found to characterize a wide variety of natural phenomena including disasters such as earthquakes, floods and forest fires~\citep{Bak89, Turcotte98, Newman05}, social behavior or organization such the distribution of city sizes, the number of citations for scientific papers, the number of participants in strikes, and the frequency of words in language~\citep{Zipf49, Simon55, Newman05, Biggs05}, among others. As a reflection of their apparent ubiquity, but somewhat pejoratively, it has even been said that such power-law statistics seem ``more normal than normal''~\citep{Willinger05}.

In this paper, we extend Richardson's program of study to the most topical kind of conflict: terrorism. Specifically, we empirically study the distributional nature of the frequency and severity of terrorist events worldwide since 1968. Although terrorism as a political tool has a long history~\citep{Congleton02, Enders06}, it is only in the modern era that small groups of so-motivated individuals have had access to extremely destructive weapons~\citep{Shubik97, FBI99}. Access to such weapons has resulted in severe terrorist events such as the \mbox{7 August 1998} car bombing in Nairobi, Kenya which injured or killed over $5200$, and the more well known attack on \mbox{11 September 2001} in New York City which killed $2749$. Conventional wisdom holds that these rare-but-severe events are {\em outliers}, i.e., they are qualitatively different from the more common terrorist attacks that kill or injure only a few people. Although that impression may be true from an operational standpoint, it is false from a statistical standpoint. The frequency-severity statistics of terrorist events are scale invariant and, consequently, there is no fundamental difference between small and large events; both are consistent with a single underlying distribution. This fact indicates that there is no reason to expect that ``major'' or more severe terrorist attacks should require qualitatively different explanations than less salient forms of terrorism.

The results of our study are significant for several reasons. First, severe events have a well documented disproportional effect on the targeted society. Terrorists typically seek publicity, and the media tend to devote significantly more attention to dramatic events that cause a large number of casualties and directly affect the target audience~\citep{Wilk97, Gartner04}. When governments are uncertain about the strength of their opponents, more severe terrorist attacks can help terrorist groups signal greater resources and resolve and thereby influence a government's response to their actions~\citep{Overgaard94}. Research on the consequences of terrorism, such as its economic impact, likewise tends to find that more severe events exert a much greater impact than less severe incidents~\citep[Ch. 9]{Enders06}. For instance,~\cite{Navarro01} report dramatic declines in share prices on the New York Stock Exchange, Nasdaq, and Amex after the devastating 11 September attacks in the United States. In contrast, although financial markets fell immediately following the 7 July 2005 bombings in London, share prices quickly recovered the next day as it became clear that the bombings had not been as severe as many initially had feared.\footnote{See figures for the FTSE 100 index of the 100 largest companies listed on the London Stock Exchange at {\tt http://www.econstats.com/eqty/eq\_d\_mi\_5.htm}.} Recent examples of this non-linear relationship abound, although the tremendous reorganization of the national security apparatus in the United States following the 11 September 2001 attacks is perhaps the most notable in Western society. Second, although researchers have made efforts to develop models that predict the {\em incidence} of terrorist attacks, without also predicting the {\em severity}, these predictions provide an insufficient guide for policy, risk analysis, and recovery management. In the absence of an accurate understanding of the severity statistics of terrorism, a short-sighted but rational policy would be to assume that every attack will be severe. Later, we will show that when we adapt current models of terrorism to predict event severity, they misleadingly predict a thin tailed distribution, which would cause us to dramatically underestimate the future casualties and consequences of terrorist attacks. Clearly, we need to better understand how our models can be adapted to more accurately produce the observed patterns in the frequency-severity statistics. That is, an adequate model of terrorism should not only give us indications of where or when events are likely to occur, but also tell us how severe they are likely to be. Toward this end, we describe a toy model that can at least produce the correct severity distribution.

Past research on conflict has tended to focus on large-scale events like wars, and to characterize them dichotomously according to their incidence or absence, rather than according to their scale or severity. This tendency was recently highlighted by~\cite{Cederman03} for modeling wars and state formation, and by~\cite{Lacina06} for civil wars. Additionally accounting for an event's severity can provide significantly greater guidance to policy makers; for instance, \cite{Cioffi-Revilla91} accurately predicted the magnitude (the base ten logarithm of total combatant fatalities) of the Persian Gulf War in 1991, which could have helped in estimating the political consequences of the war.

As mentioned above, research on terrorism has also tended to focus on incidence, rather than severity. Recently, however, two of the authors of this study demonstrated for the first time that the relationship between the frequency and severity of terrorist events exhibits the surprising and robust feature of scale invariance~\citep{Clauset05}, just as Richardson showed for wars. In a subsequent study,~\cite{Johnson05} considered data for fatal attacks or clashes in the guerilla conflicts of Colombia and Iraq, suggesting that these too exhibit scale invariance. Additionally, they claim that the time-varying behavior of these two distributions are trending toward a common power law with parameter $\alpha = 2.5$ -- a value they note as being similar to the one reported by~\cite{Clauset05} for terrorist events in economically underdeveloped nations. Johnson et al. then adapted a dynamic equilibrium model of herding behavior on the stock market to explain the patterns they observed for these guerilla conflicts. From this model, they conjecture that the conflicts of Iraq, Colombia, Afghanistan, Casamance (Senegal), Indonesia, Israel, Northern Ireland and global terrorism are all converging to a universal distribution with exactly this value of $\alpha$~\citep{Johnson06}. We will briefly revisit this idea in a later section. Finally, the recent work of~\cite{Bogen06} also considers the severity of terrorist attacks primarily via aggregate figures to assess whether there has been an increase in the severity of terrorism over time, and to forecast mortality due to terrorism.

This articles makes three main contributions. First, we make explicit the utility of using a power-law model of the severity statistics of terrorist attacks, and demonstrate the robust empirical fact that these frequency-severity statistics are scale invariant. Second, we demonstrate that distributional analyses of terrorism data can shed considerable light on the subject by revealing new relationships and patterns. And third, we show that, when adapted to predict event severity, existing models of terrorism incidence fail to produce the observed heavy-tail in the severity statistics of terrorism, and that new models are needed in order to connect our existing knowledge about what factors promote or discourage terrorism with our new results on the severity statistics.

\section{Power laws: a brief primer}
Before plunging into our analysis, and for the benefit of readers who may be unfamiliar with the topic, we will briefly consider the topics of heavy-tailed statistics and power-law distributions. What distinguishes a power-law distributions from the more familiar normal distribution is its {\em heavy tail}, i.e., in a power law, there is a non-trivial amount of weight far from the distribution's center. This feature, in turn, implies that events orders of magnitude larger (or smaller) than the mean are relatively common. The latter point is particularly true when compared to a normal distribution, where essentially no weight is far from the mean. Although there are many distributions that exhibit heavy tails, the power law is a particularly special case, being identifiable by a straight line with slope $\alpha$ on doubly-logarithmic axes\footnote{A straight line on doubly-logarithmic axes is a necessary, but not sufficient condition for a distribution to be a power law; for example, when we have only a small number of observations from an exponentially distributed variable, it can appear roughly straight on double-logarithmic axes.}, and which appears widely in physics. The power law has the particular form in which multiplication of the argument, e.g., by a factor of $2$, results in a proportional division of the frequency, e.g., by a factor of $4$, and the ratio of these values is given by the ``scaling parameter'' $alpha$. Because this relationship holds for all values of the power law, the distribution is said to ``scale'', which implies that there is no qualitative difference between large and small events.

\begin{figure}[t]
\begin{center}
\includegraphics[scale=0.55]{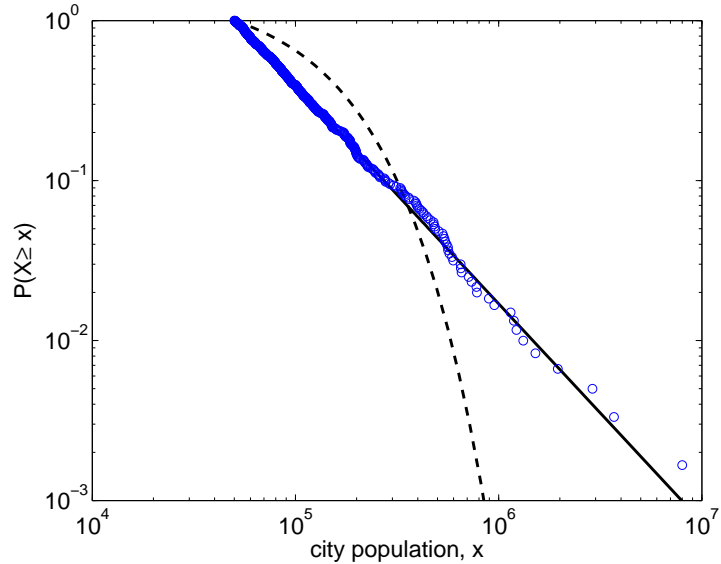}
\end{center}
\caption{The complementary cumulative distribution function (ccdf) $P(X\geq x)$ of the population $x$ of the 600 largest cities in the United States, i.e., those with $x\geq50~000$, based on data from the 2000 Census. The solid black line shows the power-law behavior that the distribution closely follows, with scaling exponent $\alpha=2.36(6)$, while the dashed black line shows a truncated normal distribution with the same sample mean.}
\label{fig:cities}
\end{figure}

Power-law distributed quantities are actually quite common, although we often do not think of them as being that way. Consider, for instance, the populations of the 600 largest cities in the United States (from the 2000 Census). With the average population being only $\langle x \rangle =165~719$, metropolises like New York City and Los Angles would seem to be clear ``outliers'' relative to this value. The first clue that this distribution is poorly explained by a truncated normal distribution is that the sample standard deviation $\sigma = 410~730$ is significantly larger than the sample mean. Indeed, if we model the data in this way, we would expect to see $1.8$ times fewer cities at least as large as Albuquerque, at $448~607$, than we actually do. Further, because it is more than a dozen standard deviations from the mean, we would never expect to see a city as large as New York City, with a population of $8~008~278$; for a sample this size, the largest city we would expect to see is Indianapolis, at $781~870$. Figure~\ref{fig:cities} shows the actual distribution, plotted on doubly-logarithmic axes, as its complementary cumulative distribution function (ccdf) $P(X\geq x)$, which is the standard way of visualizing this kind of data.\footnote{The ccdf is preferable to the probability distribution function (p.d.f) as the latter is significantly noisier in the upper tail, exactly where subtle variations in behavior can be concealed. If a distribution scales, it will continue to do so on the ccdf} The scaling behavior of this distribution is quite clear, and a power-law model (black line) of its shape is in strong agreement with the data. In contrast, the truncated normal model is a terrible fit.

As a more whimsical second example, consider a world where the heights of Americans were distributed as a power law, with approximately the same average as the true distribution (which is convincingly normal when certain exogenous factors are controlled). In this case, we would expect nearly $60~000$ individuals to be as tall as the tallest adult male on record, at $2.72$ meters. Further, we would expect ridiculous facts such as $10~000$ individuals being as tall as an adult male giraffe, one individual as tall as the Empire State Building ($381$ meters), and $180$ million diminutive individuals standing a mere $17$ cm tall. In fact, this same analogy was recently used to describe the counter-intuitive nature of the extreme inequality in the wealth distribution in the United States~\citep{Crook06}, whose upper tail is also distributed according to a power law.

Although much more could be said about power laws, we hope that the curious reader takes away a few basic facts from this diversion. First, heavy-tailed distributions do not conform to our expectations of a linear, or normally distributed, world. As such, the average value of a power law is not representative of the entire distribution, and events orders of magnitude larger than the mean are, in fact, relatively common. Second, the scaling property of power laws implies that, at least statistically, there is no qualitative difference between small, medium and extremely large events, as they are all succinctly described by a very simple statistical relationship. Readers who would like more information about power laws should refer to the extensive review by~\cite{Newman05}. With these ideas in hand, we can begin our analysis of the severity statistics of terrorism.

\section{Data sources for terrorist events}
Many organizations track terrorist events worldwide, but few provide their data in a form amenable to scientific analysis. The most popular source of information on terrorist events in the political science literature is the ITERATE data set~\citep{Mickolus04}, which focuses exclusively on transnational terrorist events involving actors from at least two countries. In principle, however, and from the standpoint of frequency and severity statistics, we see no reason to restrict our analysis to transnational events. Instead, we use the data contained in the~\citet[MIPT]{MIPT} database, which largely overlaps with the ITERATE data, but also includes fully domestic terrorist events since at least 1998. We note, however, that our analyses can easily be applied to the portion of the ITERATE data that reports event severity, and indeed, doing so yields evidence similar to that which we present here. Thus, without loss of generality and except where noted, we will focus exclusively on the MIPT data for the remainder of this article. The MIPT database is itself the compilation of the RAND Terrorism Chronology 1968-1997, the RAND-MIPT Terrorism Incident database (1998-Present), the Terrorism Indictment database (University of Arkansas \& University of Oklahoma), and DFI International's research on terrorist organizations.

By 18 June 2006, the MIPT database contained records for over $28~445$ terrorist events in more than $5000$ cities across $187$ countries worldwide since 1968. Although alternative definitions for terrorism exist, the MIPT database uses a relatively standard one that may be summarized as any violent act intended to create fear for political purposes. Each entry in the database is quite narrow: it is an attack on a single target in a single location (city) on a single day. For example, the Al Qaeda attacks in the United States on \mbox{11 September 2001} appear as three events in the database, one for each of the locations: New York City, Washington D.C. and Shanksville, Pennsylvania. Each record includes the date, target, city (if applicable), country, type of weapon used, terrorist group responsible (if known), number of deaths (if known), number of injuries (if known), a brief description of the attack and the source of the information.

Of the nearly thirty thousand recorded events, $10~878$ of them resulted in at least one person being injured or killed, and we restrict our analyses to these events as they appear to be the least susceptible to any reporting bias. Further, it is a reasonable assumption that the largest events, due their severity both in terms of casualties and political repercussions, will have the most accurate casualty estimates. Finally, if there is a systemic bias in the form of a proportionally small under- or over-estimate of event severity, it will have only a small effect the results of our statistical analysis and will not change the core result of scale invariance -- as with Richardson's study of the severity of wars, simply obtaining the correct order of magnitude of an event reveals much of the basic scaling behavior. Throughout the remainder of the paper, we take the {\em severity} of an event to be either the number of injuries, the number of deaths, or their sum (total casualties), where the severity is always at least one. Unless otherwise noted, we focus exclusively on the statistics of these values.

\begin{figure}[t]
\begin{center}
\includegraphics[scale=0.55]{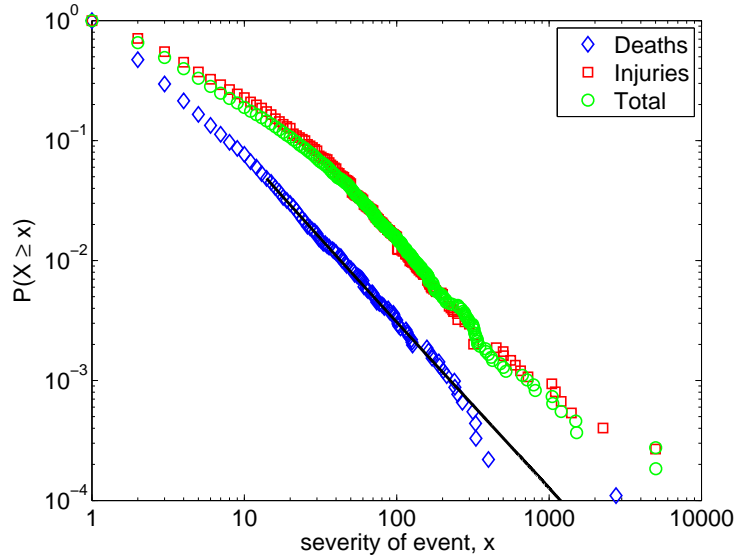}
\end{center}
\caption{The frequency-severity distributions $P(X\geq x)$ of
attacks worldwide since 1968 by injuries, deaths and
their sum. The solid line indicates the power-law scaling found by
the maximum likelihood method. Details of fits for these
distributions are given in Table 1. }
\label{fig:terrorism}
\end{figure}

\section{Frequency-severity distributions for attacks since 1968}
\label{sec:fulldistribution}
Collecting all events since 1968 as a histogram of severities, we show their complementary cumulative distribution functions (ccdfs) $P(X\geq x)$ in Figure~\ref{fig:terrorism}. The regular scaling in the upper tails of these distributions immediately demonstrates that events orders of magnitude larger than the average event size are not outliers, but are instead in concordance with a global pattern in the frequency statistics of terrorist attacks. Significantly, the scaling exists despite large structural and political changes in the international system such as the fall of communism, variations in the type of weapon used, recent developments in technology, the demise of individual terrorist organizations, and the geographic distribution of events themselves. In subsequent sections, we will examine the robustness of the scale invariance property to both categorical and temporal analysis.

If we make the idealization that events are independent and identically distributed (iid), we may model the distribution as a power law with some exponent $\alpha$, where the scaling behavior holds only for values at least some lower-bound $x_{\min}$. Obviously, significant correlations exist between many terrorist events, and such an idealization is made only for the purpose of doing a distributional analysis. Using the method of maximum likelihood, we estimate two parameters of the power-law model from the data (details of our statistical methodology are discussed in the Appendix). Models found in this way for the full distributions described above are summarized in Table 1. Using the Kolmogorov-Smirnov goodness-of-fit test, we find that these simple iid models are a surprisingly good representation of the death and total severity distributions (both $p_{\rm KS}>0.9$), although a more marginal representation of the injuries distribution ($p_{\rm KS}>0.4$).\footnote{The Kolmogorov Smirnov test evaluates whether observed data seem a plausible random sample from a given probability distribution by comparing the maximum difference between the observed and the expected cumulative distributions.}

In Section~\ref{sec:comp}, we will see that we can further decompose these distributions into their components, each of which are strongly scale invariant but with different scaling and limit parameters. As mentioned earlier, the power law is not the only distribution with a heavy tail, and although testing all such alternatives is beyond the scope of this paper, we considered another common distribution, the log-normal~\cite[see, for instance,][]{Serfling02}, and found in all cases that we may convincingly reject this model ($p_{\rm KS}<0.05$).

\begin{table}
\caption{A summary of the distributions shown in Figure~\ref{fig:terrorism}, with power-law fits from the maximum likelihood method. $N$ ($N_{\rm tail}$) depicts the number of events in the full (tail) distribution. The parenthetical value depicts the standard error of the last digit of the estimated
scaling exponent. \newline } \label{table:summary} \centering
\fbox{%
\begin{tabular}{l|ccccc|cccc}
Distribution & $N$ & $\langle x \rangle$ & $\sigma_{\rm std}$ & $x_{\rm max}$ & & $N_{\rm tail}$ & $\alpha$ & $x_{\rm min}$ & $p_{\rm KS}\geq$ \\
\hline
Injuries  & 7456 & 12.77 & 94.45 & 5000 & & 259 & 2.46(9) & 55 & 0.41 \\
Deaths   & 9101 & 4.35 & 31.58 & 2749 & & 547 & 2.38(6)  & 12 & 0.94 \\
Total       & 10878 & 11.80 & 93.46 & 5213 & & 478 & 2.48(7) & 47 & 0.99
\end{tabular}}
\end{table}

\section{Evolution of terrorism over time}
Because events in the database are annotated with their incidence date, we may write them down as a time-series and investigate the severity distribution's behavior as a function of time.\footnote{In 1998, the management of the database was transferred from the RAND Corp. to the MIPT, which resulted in several observable differences in the database records. For instance, although some purely domestic events appear prior to 1998, such as the 1995 Oklahoma City bombing, domestic events make up a significant fraction of the events entered subsequent to 1998, suggesting that the true number of events for some period directly prior to 1998 is greater than we observe in the database. Although this effect  could create problems for analyses that count incidents in a simple way, it does not effect the scale invariant shape of the frequency-severity distribution, primarily, we believe, because the large events that comprise the tail of the distribution were the least susceptible to any under-reporting bias. We shall explore this point more in the next section.} Although we are ultimately interested in the property of scale invariance over time, we first consider a simple, model-agnostic measure of the distribution's shape: the average log-severity. Sliding a window of $24$ months over the 38.5 years of event data, we compute the average log-severity (deaths) of events within each window.

\begin{figure} [t]
\begin{center}
\includegraphics[scale=0.55]{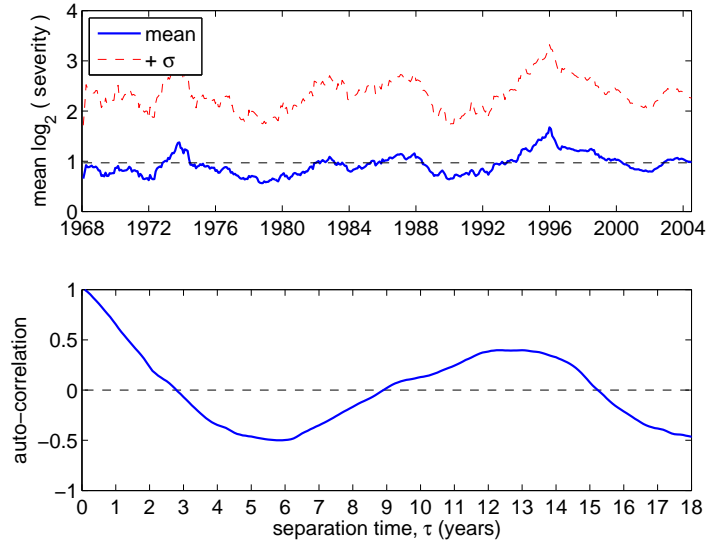}
\caption{(upper) The average log-severity (deaths) of events within a sliding window of $24$ months, for the entire 38.5 years of data. The upper dashed line indicates one standard deviation, while the other shows the average log-severity for the entire span of time. (lower) The autocorrelation function of the average log-severity, illustrating a strong periodicity in the breadth of the distribution at roughly $\tau \approx 13$ years. Similar results apply when we analyze total or injury severity, but with slight changes to the magnitude or location of the anomalous peak
in the autocorrelation function.} \label{fig:logsev}
\end{center}
\end{figure}

For highly skewed distributions, such as those we show in Figure~\ref{fig:terrorism}, the average log-severity measures the position on the independent axis of the distribution's center. The average log-severity is significantly less sensitive to variations in the length of the upper tail, which may arise from the occasional presence of rare-but-severe events, than is the average severity. The resulting time series of this measure is shown in the upper-pane of Figure~\ref{fig:logsev}, along with one standard deviation. Notably, this function is largely stable over the nearly forty years of data in the MIPT database, illustrating that the center of the distribution has not varied substantially over that time.

A closer examination of the fluctuations, however, suggests the presence of potential periodic variation. We investigate this possibility by taking the autocorrelation function (ACF) of the time series, which we show in the lower-pane. The noticeable sinusoidal shape in the ACF shows that the fluctuations do exhibit a strong degree of periodicity on the order of $\tau \approx 13$ years. If we vary the size of the window, e.g., windows between $12$ and $60$ or more months (data not shown), the location and magnitude of the peak are, in fact, quite stable. But, these features do vary slightly if we instead either examine the total or injury distributions, or truncate the time-series. As such, we conjecture that some periodicity is a natural feature of global terrorism, although we have no explanation for its origin. It has been suggested that the $\tau\approx13$ value may be related to the modal life-expectancy of the average terrorist group. However, we caution against such conclusions for now, as these aforementioned variations on our analysis can shift the peak by several years.

\section{Scale invariance over time}
Turning now to the question of scale invariance over time, we again use a sliding window of two years, but now shifted forward by one year at a time. To remain parsimonious, we make the idealization that events within each window were drawn iid from a two-parameter power-law model. After fitting such a model to each window's frequency-severity distribution, we calculate its statistical significance as a way to check the model's plausibility for that time-period.
Obviously, this assumption of no temporal correlations is quite strong, and, where appropriate, we discuss what light out analysis sheds on its accuracy. \cite{Johnson05} used a similar approach to study the time-varying distributions for the conflicts in Colombia and Iraq, but did not consider the accuracy of their models' fit or give any measure of their statistical significance.

\begin{figure} [t]
\begin{center}
\begin{tabular}{cc}
\includegraphics[scale=0.36]{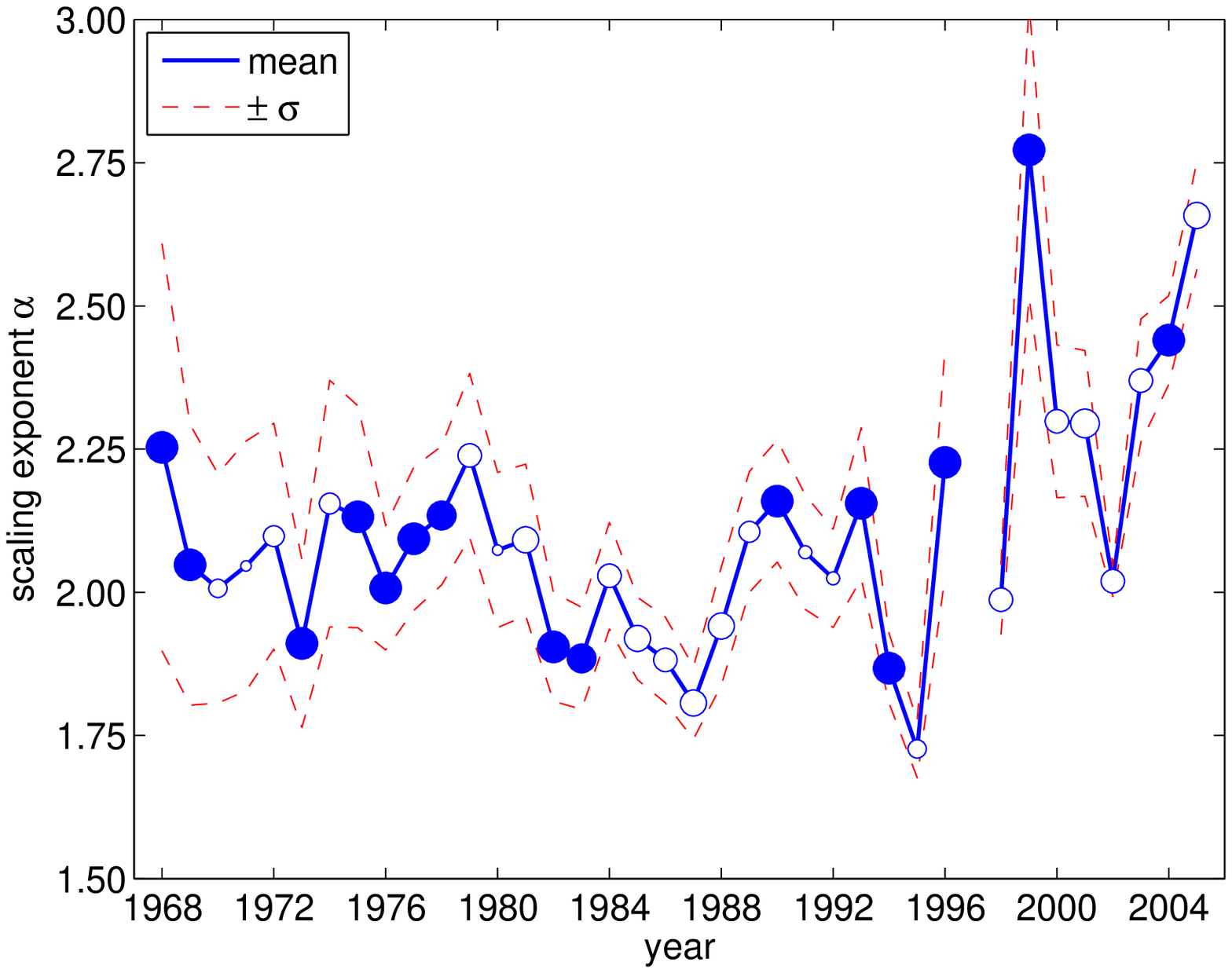} &
\includegraphics[scale=0.36]{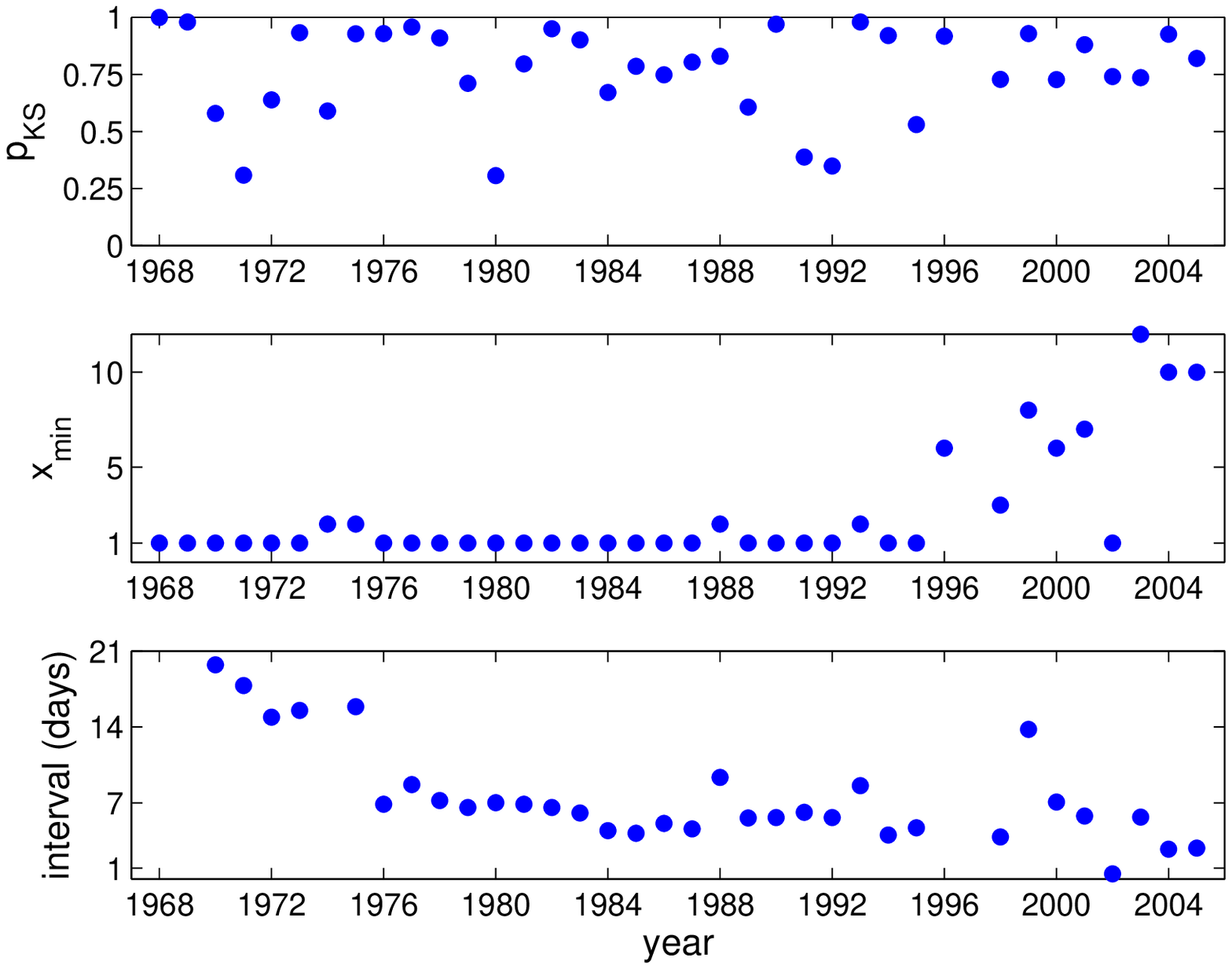} \\
(a) & (b)
\end{tabular}
\caption{Results of fitting a simple power-law model to the tail of the severity (deaths) distribution of terrorist events within discrete periods of time since 1968. We divide time into two year periods, sliding the window forward one year at a time; similar results apply for larger windows. (a) The average scaling exponent $\alpha$ for each two-year period, with circle size being proportional to the statistical significance value $p_{\rm KS}$; solid circles indicate $p>0.9$. We omit the data point for 1997 as it spans the transition of the database management. (b) Three panes showing aspects of the models; (top) significance values computed from a one-sided KS test showing that most models do not achieve high significance, (middle) the estimated $x_{\rm min}$ values, and (bottom) the average inter-event interval for events in the tail, i.e., those with severity greater than $x_{\min}$.}
\label{fig:timeseries}
\end{center}
\end{figure}

In Figure~\ref{fig:timeseries}a we show the estimated scaling parameters $\alpha$ for each time period. For the first 30 years of data, the scaling parameter appears to fluctuate around $\alpha\approx 2$, which suggests that the scaling behavior was relatively stable over this period. Subsequent to 1998, when a larger number of domestic events were incorporated into the database, the scaling parameter shifts upward to $\alpha\approx 2.5$, but again shows no consistent trend in any direction. This shift, taken with the apparent stability of the scaling behavior over time, suggests that the absence of domestic events before 1998 may have biased those distributions toward more shallow scaling, i.e., before 1998, larger events appear to be more common.

Although many ($41\%$) of these iid power-law models appear to match the distribution of severities quite well ($p_{\rm KS}>0.9$), nearly half ($49\%$) achieve only a middling level of statistical significance ($0.5< p_{\rm KS} \leq 0.9$; Figure~\ref{fig:timeseries}b upper). That is, there are significant temporal correlations within the time series, or perhaps there are strong but temporally localized deviations from the long-term structure of the power-law distribution, that cause our simple model to yield a poor fit at these times. Either case is unsurprising for this kind of real-world data. An interesting line of future inquiry would be a close study of the tail events' political context, which may reveal the origin of their correlations and explain when temporally local deviations from the long-term behavior occurred.

Further, we observe that the frequency of the most severe events, i.e., events in the upper tail of the distribution, has not changed much over the past 30 years. In Figure~\ref{fig:timeseries}b (lower pane), we plot the reciprocal of those frequencies, the mean inter-event intervals, for each two-year period. Notably, from 1977--1997, the inter-event interval for extreme events averaged $6.9\pm3.7$ days, while from 1998--2006, it averaged $5.3\pm4.0$ days.
Although this result may appear to contradict recent official
reports that the frequency of terrorist attacks worldwide has
increased dramatically in the past few decades~\citep{State04}, or that the frequency of ``major'' events has decreased, it
does not. Instead, the situation is slightly more complicated: our
analysis suggests that the changes in event frequencies have not
been evenly distributed with respect to their severity, but rather
that less severe attacks are now relatively more frequent, while
the frequency of ``major'' or tail-events has remained unchanged.
This behavior is directly observable as the upward movement of the lower-bound on the scaling region in recent years, precisely when attacks overall are thought to be more frequent (Figure~\ref{fig:timeseries}b, middle pane).

Taking the above results together with those of the average log-severity time-series (Figure~\ref{fig:logsev}) in the previous section, we can reasonably conclude that the dominant features of the frequency-severity statistics of terrorism have not changed substantially over the past 38.5 years. That is, had some fundamental characteristic of terrorism changed in the recent past, as we might imagine given recent political events, the frequency-severity distribution would not display the degree of stability we observe in these statistical experiments.

\section{Variation in scale invariance by target-country industrialization}
\begin{figure}[t]
\begin{center}
\begin{tabular*}{16.5cm}{cc}
\includegraphics[scale=0.36]{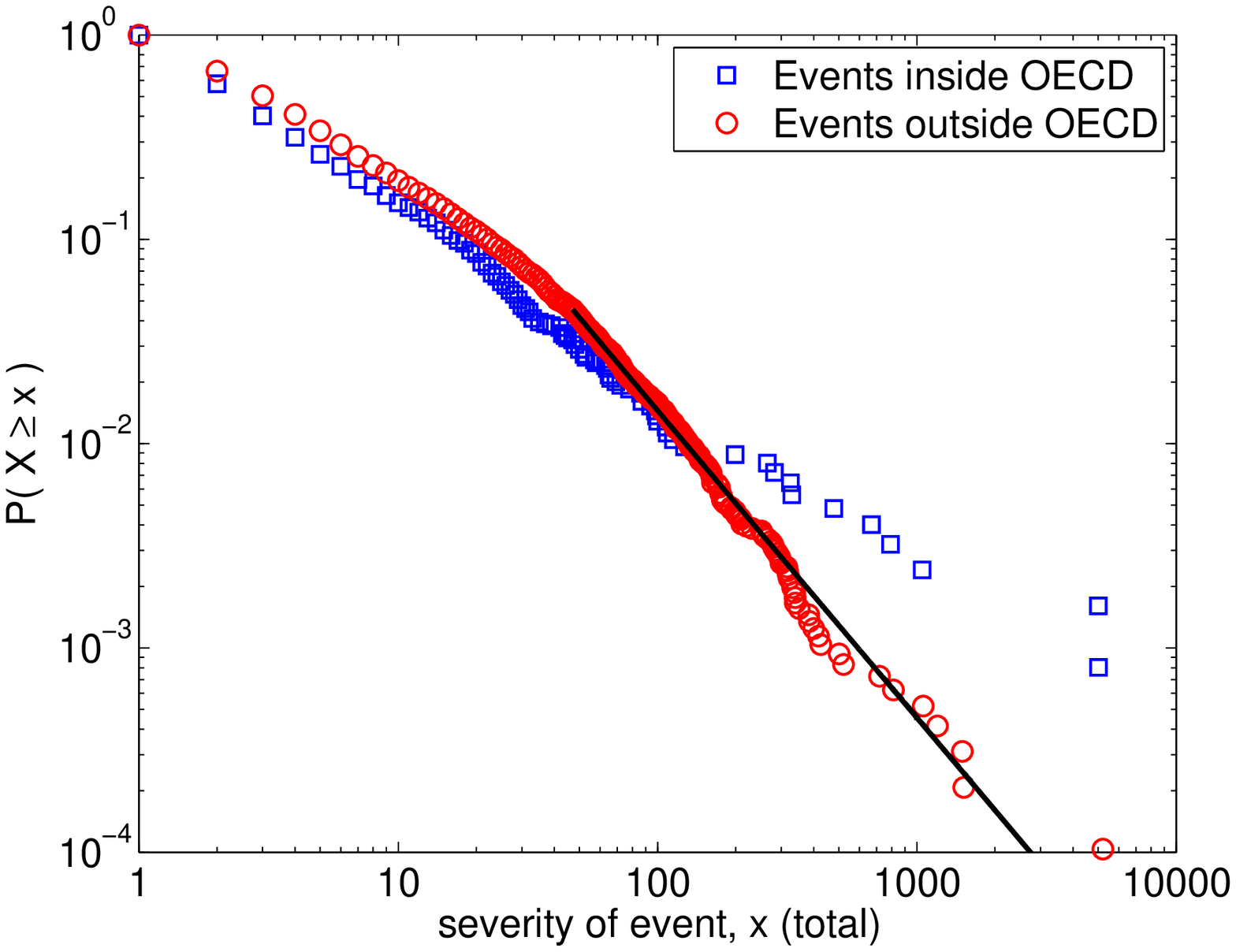} &
\includegraphics[scale=0.36]{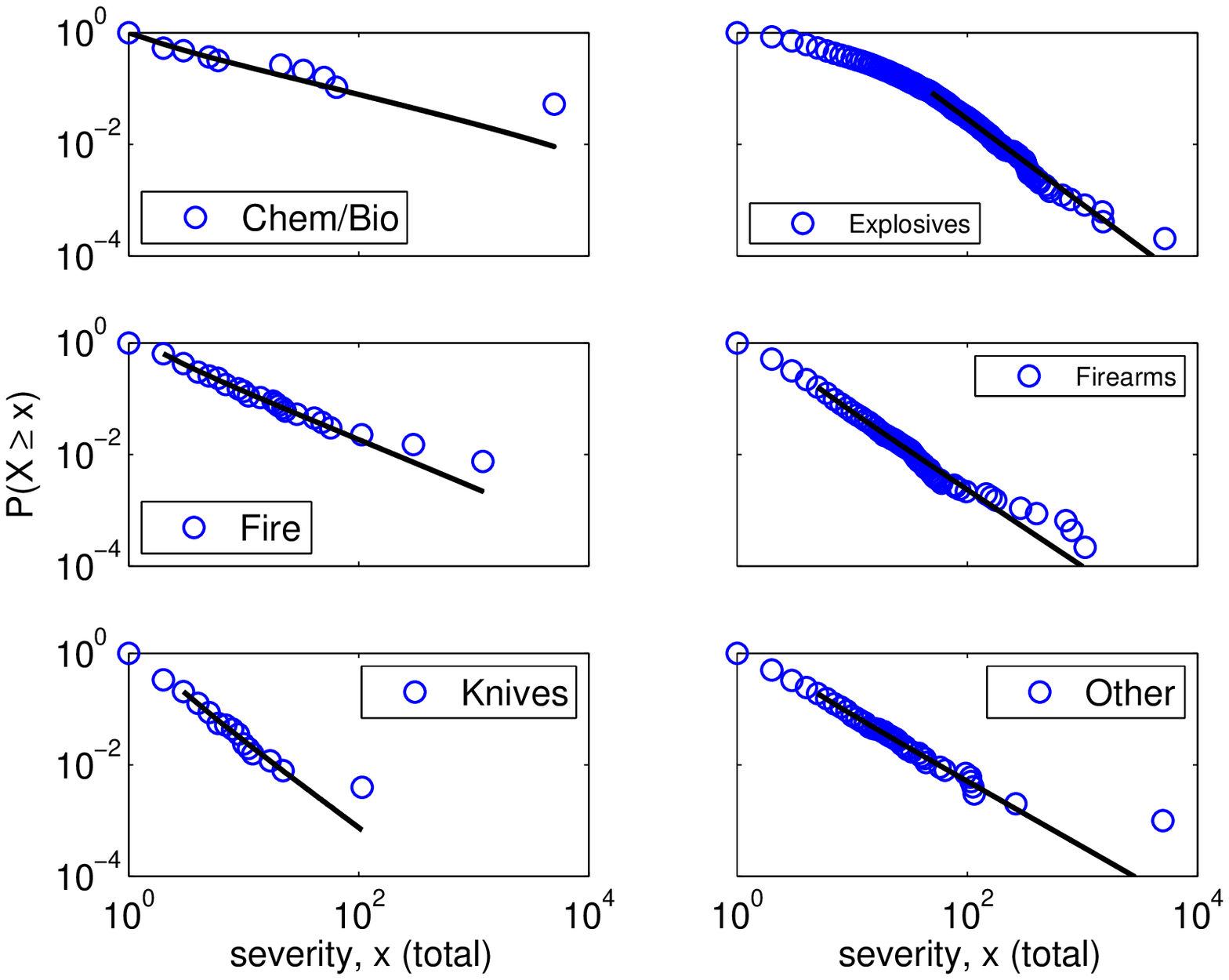} \\
(a) & (b)
\end{tabular*}
\end{center}
\caption{(a) The frequency-severity (total) distributions $P(X\geq x)$ of attacks worldwide between February 1968 and June 2006, divided among nations inside and outside of the OECD. (b) Total-severity distributions for six weapon types: chemical or biological agents ($0.2\%$ of events), explosives ($44.8\%$, includes remotely detonated devices), fire ($1.2\%$), firearms ($42.3\%$), knives ($2.3\%$) and other ($9.2\%$; includes unconventional and unknown weapon types). For both figures, solid lines indicate the fits, described in Table 2, found by the maximum likelihood method. }
\label{fig:weapons}
\end{figure}

Returning to the full distributions, we now consider the impact of industrialization on the frequency-severity statistics -- given that each attack is executed within a specific country, we may ask whether there is a significant difference in the scaling behaviors of events within industrialized and non-industrialized countries. Toward this end, we divide the events since 1968 into those that occurred within the 30 Organization for Economic Co-operation and Development (OECD) nations (1244 events, or $11\%$), and those that occurred throughout the rest of the world ($9634$ events, or $89\%$). We plot the corresponding total severity distributions in Figure~\ref{fig:weapons}a, and give their summary statistics in Table 2.

Most notably, we find substantial differences in the scaling of the two distributions, where industrialized-nation events scale as $\alpha_{\rm OECD}=2.02(9)$ while non-industrialized-nation events scale more steeply, as $\alpha_{\rm non-OECD}=2.51(7)$. That is, while events have been, to date, less likely to occur within the major industrialized nations, when they do, they tend to be more severe than in non-industrialized nations. Although this distinction is plausibly the result of technological differences, i.e., industrialization itself makes possible more severe events, it may also arise because industrialized nations are targeted by more severe attacks for political reasons. For instance, the OECD events are not uniformly distributed over the 30 OECD nations, but are disproportionately located in eight states: Turkey (335 events), France (201), Spain (109), Germany (98), the United States of America (93), Greece (76), Italy (73) and the United Kingdom (62). These eight account for $84.2\%$ (1047) of all such events, and $141$ of those are tail events, i.e., their total severity is at least $x_{\min}=13$. These eight nations account for $89.2\%$ of the most severe events, suggesting that industrialization alone is a weak explanation of the location of severe attacks, and that political factors must be important.

\section{Variation in scale invariance by weapon type}
\label{sec:comp} As our final characterization of the frequency-severity distribution's scale-invariance, we consider the connection between technology, represented by the type of weapon used in an attack, and the severity of the event. Figure~\ref{fig:weapons}b shows the total severity distributions for chemical or biological weapons, explosives (including remotely detonated devices), fire, firearms, knives and a catch-all category ``other'' (which also includes unconventional\footnote{$~$The attacks of \mbox{11 September 2001} are considered unconventional.} and unknown weapons). We find that these component distributions themselves exhibit scale invariance, each with a unique exponent $\alpha$ and lower limit of the power-law scaling $x_{\rm min}$. However, for the chemical or biological weapons, and the explosives distributions, we must make a few caveats. In the former case, the sparsity of the data reduces the statistical power of our fit, and, as discussed by~\cite{Bogen06}, the severity of the largest such event, the 1998 sarin gas attack in Tokyo, is erroneously high. For the latter distribution, another phenomenon must govern the shape of the lower tail, and we investigate its causes below. Table 2 summarizes the distributions and their power-law models.

By partitioning events by weapon-type, we now see that the origin of the bending in the lower tail of the injury and total severity distributions (Figure~\ref{fig:terrorism}a) are primarily due to explosive attacks, i.e., there is something about attacks utilizing explosives that makes them significantly more likely to injure a moderate or large number of people than other kinds of weapons. However, this property fails for larger events, and the regular scaling resumes in the upper tail. In contrast, we see no such change in the scaling behavior in the lower tail for other weapons -- this demonstrates that the property of scale invariance is largely independent of the choice of weapon. Further, by partitioning events according to their weapon type, we retain high estimates of statistical significance ($p_{\rm KS}>0.9$).

\begin{table}
\caption{A summary of the distributions shown in Figure~\ref{fig:weapons}, with power-law fits from the maximum likelihood method. $N$ ($N_{\rm tail}$) depicts the number of events in the full (tail) distribution. The parenthetical value depicts the standard error of the last digit of the estimated scaling exponent. As described in the text, the statistical significance of the explosives distribution model increases to $p_{\rm KS}\geq0.82$ when we control for suicide explosive
attacks. \newline } \label{table:wsummary} \centering
\fbox{%
\begin{tabular}{l|ccccc|cccc}
Distribution & $N$ & $\langle x \rangle$ & $\sigma_{\rm std}$ & $x_{\rm max}$ & & $N_{\rm tail}$ & $\alpha$ & $x_{\rm min}$ & $p_{\rm KS}\geq$ \\
\hline
OECD           & 1244 & 17.65 & 206.28 & 5012 & & 158 & 2.02(9) & 13 & 0.61 \\
Non-OECD  & 9634 & 11.04 & 66.09 & 5213 & & 438 & 2.51(7) & 47 & 0.84 \\
\hline
Chem/Bio     & 19 & 274.11 & 1147.48 & 5012 & & 19 & 1.5(2) & 1 & 0.89 \\
Explosives   & 4869 & 18.93 & 90.61 & 5213 & & 412 & 2.52(7) & 49 & 0.60 \\
Fire                & 133 & 16.79 & 107.14 & 1200 & & 85 & 1.9(1) & 2 & 0.99 \\
Firearms       & 4603 & 4.09 & 24.52 & 1058 & & 744 & 2.37(5) & 5 & 0.92 \\
Knives           & 254 & 2.43 & 7.01 & 107 & & 52 & 2.6(2) & 3 & 0.99 \\
Other             & 1000 & 9.30 & 158.79 & 5010 & & 189 & 2.17(9) & 5 & 0.99
\end{tabular}}
\end{table}%

What property of explosives attacks can explain the large displacement of the upper tail in that distribution? \cite{Pape03} demonstrated, through a careful analysis of all suicide attacks between 1980 and 2001, that suicide attacks cause significantly more deaths than non-suicide attacks on average, being $13$ and $1$, respectively. Similarly, for our data set, the average total severity for suicide attacks using explosives is $41.11$, while non-suicide attacks have an average total severity of $14.41$. Controlling for these attacks (692 events, or $12.9\%$) does not significantly change the curvature of the lower-tail in the explosives distribution. It does, however, improve the statistical significance of our best-fit model to the upper tail ($\alpha=2.55(9)$, $x_{\min}=47$, $p_{\rm KS}\geq 0.82$), suggesting that the severity of suicide explosives attacks deviates strongly from the general scaling behavior, and further that such attacks are not the source of the lower-tail's curvature. Conditioning on additional factors, either singly or jointly, such as the target, tactic or geographic region, can reduce the curvature in the lower-tail to varying degrees, but can never eliminate it (results not shown).

By analyzing the sequence of events, however, we find evidence that the curvature is at least partially a temporal phenomenon. When we divide events into the four decades beginning with 1968, 1978, 1988 and 1998, we see that the displacement of the upper tail $x_{\min}$ increases over time, ranging from between 2--20 for the first three decades, to 49 for the most recent decade. Indeed, because most of the explosives events in the database occurred recently (3034 non-suicide events, or $72.6\%$), the scaling behavior of this decade dominates the corresponding distribution in Figure~\ref{fig:weapons}b. Separating the data by time, however, yields more statistically significant models, with $p_{\rm KS}\geq0.8$ for the latter three decades, and progressively more curvature in the lower tail over time. Thus, we cannot wholly attribute the curvature to the inclusion of domestic events in more recent years, although certainly it is largest then. Rather, its behavior may be a function of changes in the explosives technology used in terrorist attacks over the past 40 years. The validation of this hypothesis, however, is beyond the scope of the current study, and we leave it for future work.

\section{A regression model for the severity of terrorist events}
There is an extensive literature on what factors promote terrorism and make governments more likely to become targets of terrorism. We refer to~\cite{Reich90},~\cite{Pape03}, and~\cite{Sandler05} for overviews of existing studies of terrorism. Notably, however, existing studies say nothing about the frequency-severity distribution of events, and empirical research on terrorism has tended to focus on predicting terrorist incidence. In this section, we consider to what extent models proposed to predict the incidence of terrorism data can account for the severity of terrorism, and to what extent they can reproduce the observed frequency-severity distribution.

\begin{table}
\caption{Coefficients for a negative binomial regression model on
terrorist event incidence, after~\cite{Li05}, and its ability to
predict observed severity statistics; parenthetical entries give
robust standard errors. \newline } \label{table:results}
\centering
\fbox{%
\begin{tabular}{lcccc}
\hline
    & \mbox{\rule[-0.2cm]{0cm}{0.7cm} (1)} & (2) & (3) & (4) \\
{\textbf{Variable}} & \shortstack{No. attacks \\ (ITERATE)}
    & \shortstack{No. attacks \\ (MIPT)}
    & \shortstack{Deaths by \\ event}
    & \shortstack{Deaths by \\ country-year} \\
 &  &  &  &  \\ \hline
Govt constraint & \mbox{\rule[-0.2cm]{0cm}{0.7cm} 0.061} & 0.102 & -0.013 & 0.046 \\
\vspace{4pt} & \begin{footnotesize}(0.023)\end{footnotesize}
    & \begin{footnotesize}(0.030)\end{footnotesize}
    & \begin{footnotesize}(0.013)\end{footnotesize}
    & \begin{footnotesize}(0.038)\end{footnotesize}\\
Democratic participation  & -0.009 & -0.007 & -0.001 & -0.011 \\
\vspace{4pt} & \begin{footnotesize}(0.004)\end{footnotesize}
    & \begin{footnotesize}(0.006)\end{footnotesize}
    & \begin{footnotesize}(0.003)\end{footnotesize}
    & \begin{footnotesize}(0.007)\end{footnotesize} \\
Income inequality & 0.001 & -0.001 & 0.003 & -0.002 \\
\vspace{4pt} & \begin{footnotesize}(0.014)\end{footnotesize} & \begin{footnotesize}(0.016)\end{footnotesize} & \begin{footnotesize}(0.007)\end{footnotesize} & \begin{footnotesize}(0.021)\end{footnotesize} \\
Per capita income & -0.177 & -0.161 & 0.008 & -0.222 \\
\vspace{4pt} & \begin{footnotesize}(0.11)\end{footnotesize} & \begin{footnotesize}(0.14)\end{footnotesize} & \begin{footnotesize}(0.047)\end{footnotesize} & \begin{footnotesize}(0.15)\end{footnotesize} \\
Regime durability & -0.076 & -0.109 & 0.039 & 0.010 \\
\vspace{4pt} & \begin{footnotesize}(0.047)\end{footnotesize} & \begin{footnotesize}(0.060)\end{footnotesize} & \begin{footnotesize}(0.024)\end{footnotesize} & \begin{footnotesize}(0.067)\end{footnotesize} \\
Size & 0.118 & 0.0494 & -0.014 & -0.001 \\
\vspace{4pt} & \begin{footnotesize}(0.044)\end{footnotesize} & \begin{footnotesize}(0.054)\end{footnotesize} & \begin{footnotesize}(0.015)\end{footnotesize} & \begin{footnotesize}(0.079)\end{footnotesize} \\
Govt capability & 0.275 & 0.189 & -0.018 & 0.072 \\
\vspace{4pt} & \begin{footnotesize}(0.14)\end{footnotesize} & \begin{footnotesize}(0.18)\end{footnotesize} & \begin{footnotesize}(0.061)\end{footnotesize} & \begin{footnotesize}(0.21)\end{footnotesize} \\
Past incident & 0.547 & 0.717 & -0.009 & 0.789 \\
\vspace{4pt} & \begin{footnotesize}(0.045)\end{footnotesize} & \begin{footnotesize}(0.052)\end{footnotesize} & \begin{footnotesize}(0.024)\end{footnotesize} & \begin{footnotesize}(0.081)\end{footnotesize} \\
Post-cold war  & -0.578 & -0.253 & 0.104 & -0.036 \\
\vspace{4pt} & \begin{footnotesize}(0.097)\end{footnotesize} & \begin{footnotesize}(0.11)\end{footnotesize} & \begin{footnotesize}(0.061)\end{footnotesize} & \begin{footnotesize}(0.16)\end{footnotesize} \\
Conflict & -0.170 & -0.046 & 0.294 & 0.072 \\
\vspace{4pt} & \begin{footnotesize}(0.11)\end{footnotesize} & \begin{footnotesize}(0.13)\end{footnotesize} & \begin{footnotesize}(0.13)\end{footnotesize} & \begin{footnotesize}(0.39)\end{footnotesize} \\
Europe & 0.221 & -0.263 & -0.133 & -0.589 \\
\vspace{4pt} & \begin{footnotesize}(0.20)\end{footnotesize} & \begin{footnotesize}(0.34)\end{footnotesize} & \begin{footnotesize}(0.075)\end{footnotesize} & \begin{footnotesize}(0.49)\end{footnotesize} \\
Asia & -0.494 & -0.684 & 0.239 & -0.542 \\
\vspace{4pt} & \begin{footnotesize}(0.25)\end{footnotesize} & \begin{footnotesize}(0.28)\end{footnotesize} & \begin{footnotesize}(0.13)\end{footnotesize} & \begin{footnotesize}(0.36)\end{footnotesize} \\
America & -0.349 & -0.681 & -0.098 & -1.125 \\
\vspace{4pt} & \begin{footnotesize}(0.15)\end{footnotesize} & \begin{footnotesize}(0.23)\end{footnotesize} & \begin{footnotesize}(0.073)\end{footnotesize} & \begin{footnotesize}(0.30)\end{footnotesize} \\
Africa & -0.423 & -0.462 & 0.022 & -0.538 \\
\vspace{4pt} & \begin{footnotesize}(0.18)\end{footnotesize} & \begin{footnotesize}(0.21)\end{footnotesize} & \begin{footnotesize}(0.12)\end{footnotesize} & \begin{footnotesize}(0.31)\end{footnotesize} \\
Constant & -0.443 & 0.805 & 1.591 & 2.548 \\
\vspace{4pt} & \begin{footnotesize}(1.54)\end{footnotesize} &
    \begin{footnotesize}(1.89)\end{footnotesize} &
    \begin{footnotesize}(0.65)\end{footnotesize} &
    \begin{footnotesize}(2.63)\end{footnotesize} \\
\multicolumn{5}{c}{} \\ \hline
N & \mbox{\rule[-0.2cm]{0cm}{0.7cm} 2232} & 2232 & 1109 & 2232 \\
Log-likelihood & -3805.791 & -3300.011 & -2897.375 & -2268.129 \\
LR-$\chi^{2}$ & 1151.842 & 507.427 & 151.709 & 373.293 \\ \hline
\end{tabular}
}
\end{table}

As a recent example of empirical studies on the frequency of terrorist attacks, we use that of~\cite{Li05}. Although different studies have suggested different features to predict variation in terrorist incidents, the Li study is both careful and generally representative of the structure of cross-country comparative studies. Li empirically explores the impact of a large number of political and economic factors that have been hypothesized to make transnational terrorist incidents more or less likely, and argues that while some features of democratic institutions, such as greater executive constraints, tend to make terrorist incidents more likely, other features, such as democratic participation, are associated with fewer incidents. Model (1) in Table 3 displays the coefficient estimates for Li's original results from a negative binomial regression of the number of transnational terrorist events, with each country-year as the unit of observation. We refer to the original~\cite{Li05} article for all details on variable construction, etc.

Since our data are based on terrorist incidents that are not limited to transnational events, 
we first replicate the Li model for incidents in the MIPT data to ensure that our results are not an artifact of systematic differences between transnational-only and transnational-plus-domestic terrorist events. The coefficient estimates for the Li model applied to the number of incidents in the MIPT data shown as Model (2) in Table 3 are generally reasonably similar to the results for the original Model (1), suggesting that the model behaves similarly when applied to the two sources of data on terrorism.

Next, we examine to what extent the right-hand side covariates in the Li model allow us to predict to differences in the severity of terrorism. Model (3) in Table 3 
displays the results for a negative binomial regression of the number of deaths among the lethal events in the MIPT data. Comparing the size of the coefficient estimates to their standard errors suggest that none of these coefficients are distinguishable from 0, with the possible exception of the estimate for Europe and the post-Cold War period. In other words, none of the factors proposed by Li seem to be good predictors of the severity of terrorist events. Moreover, the proposed Li model fails to generate predictions that in any way resemble the observed variation in the number of deaths: the largest predicted number of deaths for any observation in the observed sample is less than 10, far below the actual observed maximum of 2749 (i.e., the 11 September 2001 attack on the World Trade Center).

The original Li model examines the number of incidents by country-year, and it may therefore be argued that looking only at events with casualties could understate the possible success of the model in identifying countries that are unlikely to become targets of terrorist incidents. The results for the Li model applied to the total events for all country-years, Model (4) in Table 3, however, do not lend much support to this idea. Very few of the features emphasized by Li have coefficient estimates distinguishable from 0 by conventional significance criteria, and the highest predicted number of deaths for any one country-year in the sample is still less than 16. As such, this model is clearly not able to generate the upper tail of the observed frequency-severity distribution.

\section{A toy model for scale invariance through competitive forces}
Having shown that  a representative model of terrorism incidence is a poor predictor of event severity, we now consider an alternative mechanism by which we can explain the robust statistical feature of scale invariance. As it turns out, power law distributions can arise from a wide variety of processes~\citep{Kleiber03, Mitzenmacher04, Newman05, Farmer06}. In the case of disasters such as earthquakes, floods, forest fires, strikes and wars, the model of self-organized criticality (SOC)~\citep{Bak87}, a physics model for equilibrium critical phenomena\footnote{Critical phenomena characterize a phase transition such as the evaporation of water, while an equilibrium critical phenomenon is one in which the critical state is a global attractor of system dynamics.} in spatially extended systems, appears to be the most reasonable explanation~\citep{Bak89,Turcotte98,Cederman03,Biggs05} as events themselves are inherently spatial. However, such models seem ill-suited for terrorism, where the severity of an event is not merely a function of the size of the explosion or fire. That is, the number of casualties from a terrorist attack is also a function of the density of people at the time and location of the attack, and of the particular application of its destructive power, e.g., a small explosion on an airplane can be more deadly than a large explosion on solid ground.\footnote{A trivial ``spatial'' model for the frequency-severity scale invariance would be a tight connection with size of the targeted city and the number of casualties. That is, as we saw earlier, large city populations are distributed as a power law, and we might suppose that an event's severity is proportional to the size of the target city. If target cities are chosen roughly uniformly at random, an obviously unrealistic idealization, then a power law in the frequency-severity statistics follows naturally. Tabulating population estimates for cities in our database from publicly available census data, we find that the correlation between an event's severity and the target city population is very weak, $r = 0.2(2)$ for deaths and $r = 0.2(1)$ for total severity, where the number in parentheses is the standard error from a bootstrap resampling of the correlation calculation.}

In the context of guerilla conflicts,~\citeauthor{Johnson05}~\citeyearpar{Johnson05,Johnson06} have adapted a dynamic equilibrium model of herding behavior on the stock market to produce frequency-severity distributions with exponents in the range of $1.5$ to $3.5$, depending on a parameter that is related to the rates of fragmentation and coalescence of the insurgent groups; they conjecture that the value $2.5$ is universal for all asymmetric conflict, including terrorism. Given the variation in the scaling behaviors that we measure for different aspects of terrorism (Figures~\ref{fig:terrorism},~\ref{fig:timeseries}a and~\ref{fig:weapons}a,b), this kind of universalism may be unwarranted. As an alternative explanation to the origin of the scale invariance for terrorism, we propose and analyze a simple, non-spatially extended toy model of a stochastic, competitive process between states and non-state actors~\citep{Clauset05}. The model itself is a variation of one described by~\cite{Reed02}, and can produce exponents that vary depending on the choice of model parameters -- a feature necessary to explain the different scaling behaviors for industrialization and weapon types.
Central to our model are two idealizations: that the potential severity of an event is a certain function of the amount of planning required to execute it, and that the competition between states and non-state actors is best modeled by a selection mechanism in which the probability that an event is actually executed is inversely related to the amount of planning required to execute it.

Consider a non-state actor (i.e., a terrorist) who is planning an attack. Although the severity of the event is likely to be roughly determined before planning begins, we make the
idealization that the potential severity of the event grows with time, up to some finite limit imposed perhaps by the choice of weapon (as suggested by Figure~\ref{fig:weapons}), the choice of target, or the availability of resources. If we further assume that the payoff rate on additional planning is proportional to the amount of time already invested, i.e., increasing the severity of a well-planned event is easier than for a more ad hoc event, then the potential severity of the event can be expressed as $p(t)\propto \e^{\kappa t}$, where $\kappa>0$ is a constant.

However, planned events are often prevented, aborted or executed prematurely, possibly as a result of intervention by a state. This process by which some events are carried out, while others are not, can be modeled as a selection mechanism. Assuming that the probability of a successful execution is exponentially related to the amount of time invested in its planning, perhaps because there is a small chance at each major step of the planning process that the actors will be incarcerated or killed by the state, or will abandon their efforts, we can relate the severity of a real event to the planning time of a potential event by $x \propto \e^{\lambda t}$, where $\lambda<0$ is a constant. Thus, to derive the distribution of real event severities, after the selection mechanism has filtered-out those events that never become real, we must solve the following identity from probability theory\footnote{Note that this operation is isomorphic randomly sampling the potential severity distribution.}
\[ \int p(x)\, \dx = \int p(t)\, \dt \enspace . \]
Doing so yields $p(x) \propto x^{-\alpha}$ where $\alpha = 1 - \kappa/\lambda$. Again considering the competitive nature of this process, it may be plausible that  states and actors will, through interactions  much like the co-evolution of parasites and hosts, develop roughly equal capabilities, on average, but perhaps with a slight advantage toward the state by virtue of its longevity relative to terrorist organizations, such that $|\kappa| \gtrsim |\lambda|$. In this case, we have a power law with exponent $\alpha \gtrsim 2$, in approximate agreement with much of our empirical data.

Although our toy model makes several unrealistic idealizations, its foundational assumptions fit well with the modern understanding of terrorism, and also with examples of recent attacks and foiled attempts. Whereas the plans for 11 September 2001 attacks in the United States are believed to have been underway since 1996,\footnote{On the planning for the 11 September 2001 attacks, see the summary of a documentary aired by Al-Jazeera at {\tt archives.cnn.com/2002/WORLD/meast/09/12/alqaeda.911.claim/index.html}.} subsequent attacks and attempts in the United Kingdom carried out by less organized groups and with less advance planning have failed to create a similar impact. For example, the 21 July 2005 attacks on the London Underground are now believed to have been a direct copycat effort initiated after the prior 7 July bombings. The attack was spectacularly unsuccessful: none of the four bombs' main explosive charges actually detonated, and the only reported casualty at the time was later found to have have died on an asthma attack. Even though the suspects initially managed to flee, all were later apprehended.

The competitive relationship of states and non-state actors has been explored in a variety of other contexts. \cite{Hoffman99} suggests that the state's counter-terrorism efforts serve as a selective measure, capturing or killing those actors who fail to learn from their peers' or predecessors' mistakes, leaving at-large the most successful actors to execute future attacks. \cite{Overgaard94},~\cite{Sandler03},~\cite{Sandler04}, and~\cite{Arce05}
give a similar view, arguing that the actions of states and actors are highly interdependent -- that actors typically make decisions on who, where, when or what to attack based on a careful assessment of the likelihood and impact of success, with these factors being intimately related to the decisions states make to discourage certain forms of attacks or responses. Governments make a similar calculus, although theirs is primarily reactive rather than proactive~\citep{Arce05}. Looking forward, a game theoretic approach, such as the one used by~\cite{Sandler03} to produce practical counter-terrorism policy suggestions, will likely be necessary to capture this interdependence, although presumably it will be roughly similar to the selective process we describe above. 

Obviously, the practical, geopolitical and cultural factors relevant to a specific terrorist attack are extremely complex. Although our toy model intentionally omits them, they presumably influence the values assumed by the model parameters and are essential for explaining the variety of scaling exponents we observe in the data, e.g., the different scaling exponents for OECD and non-OECD nations and for attacks perpetrated using different weapons. It may be possible to incorporate these factors by using a regression approach to instead estimate the parameter values of our toy model, rather than to directly estimate the event severity.

\section{Discussion and conclusions}
Many of the traditional analyses of trends in terrorism are comparative, descriptive, historical or institutional, and those that are statistical rely on assumptions of normality and thus treat rare-but-severe events as qualitatively different from less severe but common events~\citep{Reich90, FBI99, State04, Sandler05}. By demonstrating that Richardson's discovery of scale invariance in the frequency-severity statistics of wars extends to the severity statistics of terrorism, we show that these assumptions are fundamentally false. Our estimates of the scaling behavior for terrorism, however, differs substantially from that of the severity of wars; in the latter case, the frequency-severity distribution scales quite slowly, with $\alpha_{\rm war} = 1.80(9)$, while the distribution scales much more steeply for terrorism, $\alpha_{\rm deaths} = 2.38(6)$, indicating that severe events are relatively less common in global terrorism than in interstate warfare.

Taking Richardson's program of study on the statistics of deadly human conflicts together with the extensive results we discuss here, our previous, preliminary study of terrorism~\citep{Clauset05}, and the study by~\citeauthor{Johnson05}~\citeyearpar{Johnson05,Johnson06} of insurgent conflicts, we conjecture first that scale invariance is a {\em generic feature} of the severity distribution of all deadly human conflicts, and second that it is {\em the differences in the type of conflict} that determine the particular scaling behavior, i.e., the values of the scaling exponent $\alpha$ and the lower-limit of the scaling $x_{\min}$. Indeed, this variation is precisely what we observe when we control for attributes like the degree of economic development, and the type of weapon used in the attack. In honor of Richardson and his pioneering interest in the statistics of deadly conflict, we call our conjecture Richardson's Law. A significant open question for future work remains to determine how and why the distinguishing attributes of a conflict, such as the degree of asymmetry, the length of the campaign, and the political agenda, etc., affect the observed scaling behavior.

With regard to counter-terrorism policy, the results we describe here have several important implications. First, the robustness of the scale invariant relationship between the frequency and severity of attacks demonstrates the fact that severe events are not fundamentally different from less severe ones. As such, policies for risk analysis and contingency planning should reflect this empirical fact. Second, although severe events do occur with much greater frequency than we would expect from our traditional thin-tailed models of event severity, their incidence has also been surprisingly stable over the past 30 years (Figure~\ref{fig:timeseries}b, lower pane). This point suggests that, from an operational standpoint, and with respect to their frequency and severity, there is nothing fundamentally new about recent terrorist activities, worldwide. Third, limiting access to certain kinds of weapons and targets is clearly important, with this being particularly true for those that are inherently more likely to produce a severe event, such as high explosives, or targets like airplanes and other mass transit systems. But, severe events themselves are not only associated with one or a few weapon-types (or targets). Restricting access to some weapons and targets will likely induce the substitution of less easily restricted ones~\citep{Enders06} -- a contingency for which we should plan. Fourth, the trend we identify for explosives, i.e., that such attacks have produced progressively more casualties over time, is particularly worrying given the sheer number of explosives attacks in the recent past. Both their severity and their popularity suggest that current international regulation of explosives technology is failing to keep these weapons out of the hands of terrorists, and that current diplomacy is failing to keep terrorists from resorting to their use. And finally, although it may be tempting to draw an analogy between terrorism and natural disasters, many of which also follow power-law statistics, we caution against such an interpretation. Rather, a clear understanding of the political and socioeconomic factors that encourage terrorist activities, and an appropriate set of policies that directly target these factors, may fundamentally change the frequency-severity statistics in the future, and break the statistical robustness of the patterns we have observed to date.

In closing, the discovery that the frequency of severe terrorist attacks follows a robust empirical law opens many new questions, and points to important gaps in our current understanding of both the causes and consequences of terrorism. Although we have begun to address a few of those, such as showing that the severity of suicide attacks using explosives does not follow the same frequency-severity statistics as other forms of terrorism, many more remain. We hope to see the community of conflict researchers making greater use of these new ideas in future research on terrorism.

\section*{Acknowledgments}
\noindent A.C. and M.Y. thank Cosma Shalizi, Cristopher Moore and Raissa D'Souza for helpful conversations. K.S.G. thanks Lindsay Heger, David A. Meyer, and Quan Li. We are also grateful to the editor at JCR and two anonymous reviewers for valuable comments on a previous version of the manuscript. This work was supported in part by the National Science Foundation under grants PHY-0200909 and ITR-0324845 (A.C.), CCR-0313160 (M.Y.), and SES-0351670 (K.S.G.), and by the Santa Fe Institute (A.C.).
    
\begin{appendix}
\section*{Appendix: Statistical methodology, and the use of power laws in empirical studies}
\label{appendix:A} Because the use of power laws and other heavy-tailed distributions in the social sciences is a relatively new phenomenon, the statistical tools and their relevant characteristics may not be familiar to some readers. This appendix thus serves to both explain our statistical methodology and to give the interested reader a brief tutorial on the subject. We hope that this material illuminates a few of the subtleties involved in using power laws in real-world situations. Readers interested in still more information should additionally refer to~\cite{Newman05} and~\cite{Goldstein04}.

To begin, we note that there are two distinct kinds of power laws, a real-valued or continuous kind and a discrete kind. Although both forms have many characteristics in common, the numerical methods one employs in empirical studies can be quite different depending on whether the data are best treated as continuous or discrete. Examples of the former might be voltages on power lines, the intensity of solar flares or the magnitude of earthquakes. In cases where discrete data takes values that are quite large, they can often be safely treated as if they were continuous variables, such as for the population of US cities, books sales in the US or the net worth of Americans. In the social sciences, however, data more frequently assume integer values where the maximum value is only a few orders of magnitude larger than the minimum, i.e., the tail is heavy but rather short. Examples of this kind of data might be the number of connections per person in a social network, casualty statistics for terrorist attacks, and word frequencies in a text. If such data are treated as a continuous variable, estimates of the scaling behavior or other statistical analyses can be significantly biased.

Instead, these heavy but relatively short tails should be modeled explicitly as a discrete power law,
\begin{align*}
P(x) = x^{-\alpha} / \zeta(\alpha) \enspace,
\end{align*}
with $x$ assuming only integer values greater than zero, and $\zeta(\alpha)$ being the Riemann zeta function, the normalization constant. In what follows, we will first consider the necessity of generating a random deviate with a power-law distribution, and then consider methods for estimating power law parameters from data itself. Both sections describe the statistical methods employed in this study, and provide a brief comparison with alternative methods.

\subsection*{Generating power-law distributed data}
Statistical modeling often necessitates the generation of random deviates with a specified distribution, e.g., in simple null-models or statistical hypothesis tests. \cite{Newman05} gives a simple analytic formula, derived using the transformation method~\cite{Press92}, for converting a uniform deviate into a continuous power law deviate:
\begin{align*}
x = x_{\min}(1-r)^{-1/(\alpha-1)} \enspace,
\end{align*}
where $x$ is distributed as a real number over the interval $[x_{\min},\infty]$, and $r$ is a uniform deviate. Although it may be tempting to simply take the integer portion of each deviate $x$ in order to obtain a discrete power law, the resulting distribution will actually differ quite strongly from what is desired: such a procedure shifts a significant amount of probability mass from smaller to larger values, relative to the corresponding theoretical discrete power-law distributed deviate.

\begin{figure}[t]
\begin{center}
\begin{tabular}{cc}
\includegraphics[scale=0.36]{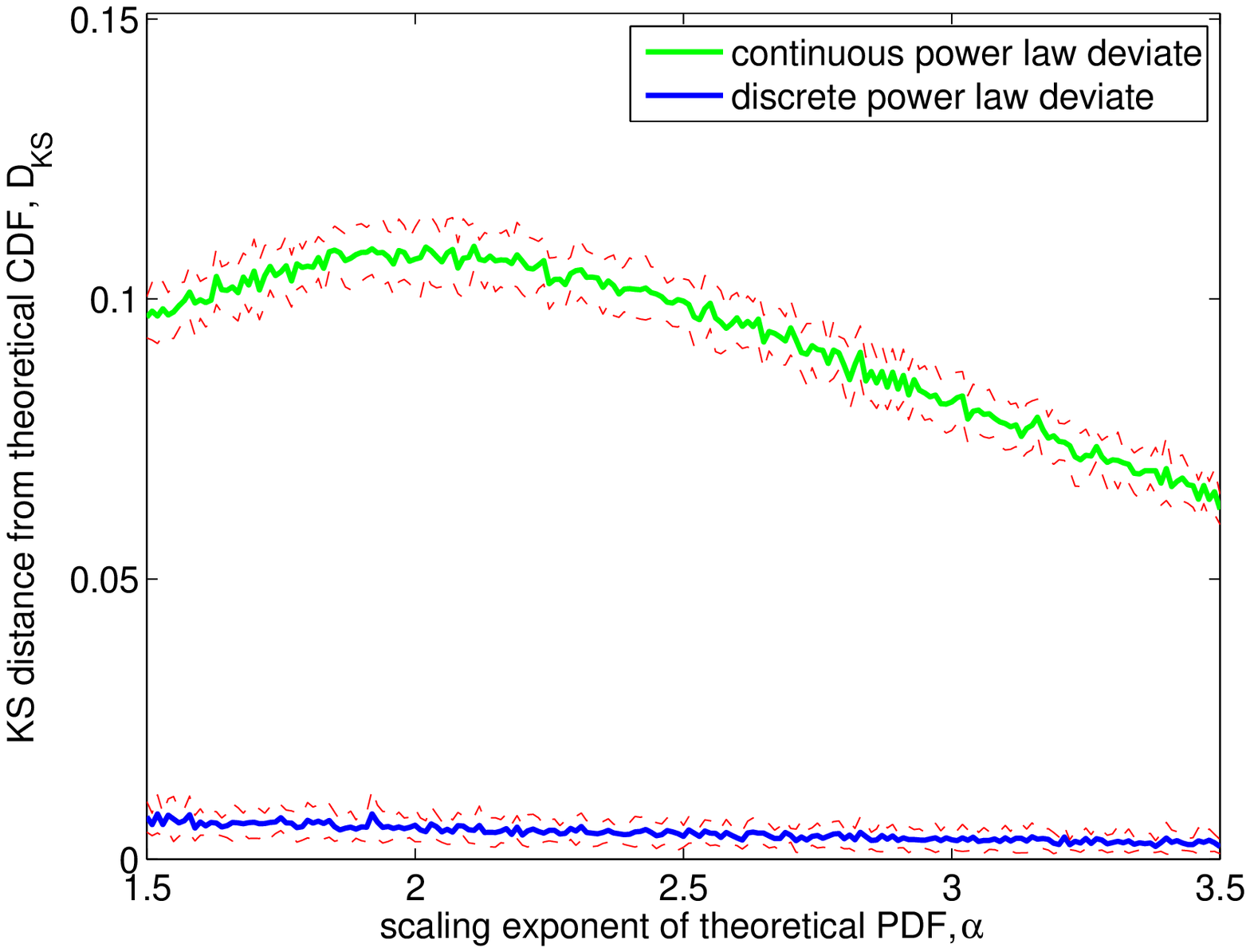} &
\includegraphics[scale=0.36]{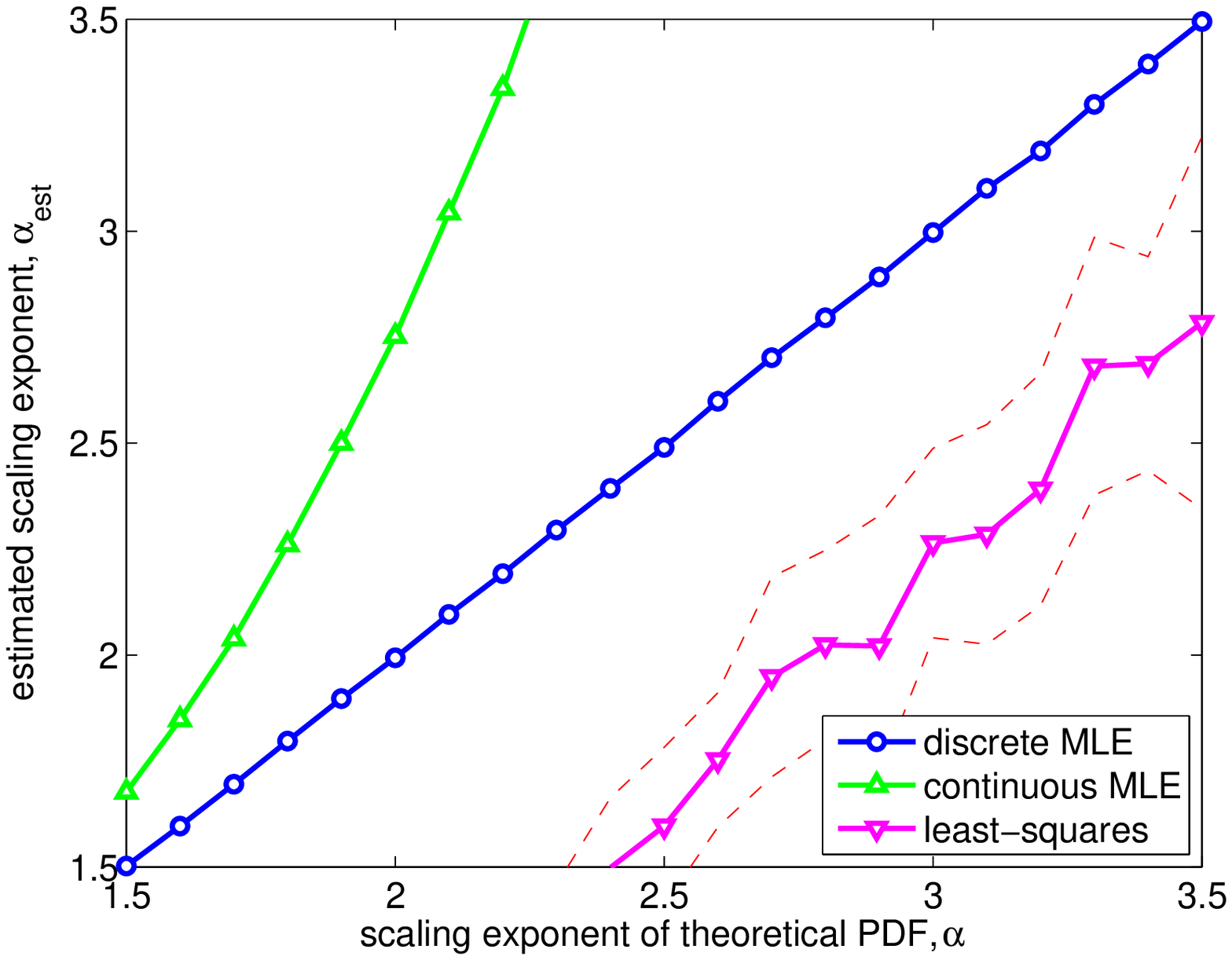} \\
(a) & (b)
\end{tabular}
\end{center}
\caption{(a) The closeness, in the sense of the Kolmogorov-Smirnov goodness-of-fit measure, of power-law distributed deviates, generated using the two methods described in the test, to the target distribution, a discrete power law with scaling parameter $\alpha=2.5$. Results are for $x_{\min}=1$ and $n=10~000$, with similar results holding for other values, although the difference decreases as $x_{\min}\rightarrow\infty$. Quite dramatically, the discrete deviate generator does a significantly better job at matching the theoretical distribution than does the continuous method discussed in the text. (b) The results of using the three methods discussed in the text for estimating the scaling parameter of discrete power-law distributed data, with parameters $x_{\min}=1$ and $n=10~000$; similar results hold for other values, although the estimates get increasingly noisy as the number of observations shrinks, and the two maximum likelihood estimators increasingly agree as $x_{\min}\rightarrow\infty$. Error bars are omitted when they are less than the size of the series symbol.} \label{fig:deviates}
\end{figure}

A more satisfying approach is to use a deviate generator specifically designed for a discrete power law. Because the discrete form does not admit a closed-form analytical solution via the transformation method like the continuous form, the generator must instead take an algorithmic approach to convert uniform deviates via the inverse cumulative density function of the discrete power law. Such an approach is a standard practice, and fast algorithms exist for doing so~\citep{Press92}. To illustrate the differences between these two power-law deviate generators, we show in Figure~\ref{fig:deviates}a that the latter approach produces distributions that are significantly closer to the desired theoretical one than does the former method, and it is the latter which we use for our statistical studies in the main text.

\subsection*{Estimating scaling parameters from data}
Since Richardson first considered the scale invariance in the frequency and severity of wars, statistical methods for characterizing power laws have advanced significantly. The signature feature of a tail distribution that decays as a power law is a straight-line with slope $\alpha$ on doubly logarithmic axes. As such, a popular method of measuring the scaling exponent $\alpha$ has been by a least-squares regression on log-transformed data, i.e., one takes the log of both the dependent and independent variables, or one could bin the data into decades, and then measures the slope using a least-squares linear fit. Unfortunately, this procedure yields a biased estimate for the scaling exponent~\citep{Goldstein04}. For continuous power-law data, Newman (2005) gives an unbiased estimator based on the method of maximum likelihood; however, it too yields a biased estimate when applied to discrete data like ours. \cite{Goldstein04} studied the bias of some estimators for power-law distributed data, and, also using the method of maximum likelihood, give a transcendental equation whose solution is an unbiased estimator for discrete data. In our main study, we use a generalization of this equation as our discrete maximum likelihood estimator.

To give the reader a sense of the performance of these methods, we show in Figure~\ref{fig:deviates}b  the results of applying them to simulated data derived from the discrete generator described above. Quite clearly, the discrete maximum likelihood estimator yields highly accurate results, with the other techniques either over- or under-estimating the true scaling parameter, sometimes dramatically so. \cite{Johnson06} have also studied the accuracy of these estimators, but apparently only for data derived from the continuous deviate generator described above.

The discrete maximum likelihood estimator of Goldstein et al. assumes that the tail encompasses the entire distribution. A generalization of their formula to distributions where the tail begins at some minimum value $x_{\min}\geq1$ follows, and the value of $\alpha_{\rm ML}$ that satisfies this equation is the discrete maximum likelihood estimator:
\begin{align*}
\frac{\zeta'(\alpha,x_{\min})}{\zeta(\alpha,x_{\min})} = -\frac{1}{n}\sum_{i=1}^{n}\log x_{i} \enspace ,
\end{align*}
where the $x_{i}$ are the data in the tail, $n$ is the number of such observations, and $\zeta(\alpha,x_{\min})$ is the incomplete Riemann zeta function. If desired, the latter can be rewritten as $\zeta(\alpha) - {\rm H}_{x_{\min}}^{\alpha}$, being the difference between a zeta function and the $x_{\min}$th harmonic number of order $\alpha$. When $x_{\min}=1$, the left-hand side reduces to $\zeta'(\alpha)/\zeta(\alpha)$, the values of which can be calculated using most standard mathematical software. Alternatively, one can numerically maximize the log-likelihood function itself,
\begin{align*}
\mathcal{L}(\alpha~|~x) = -n\log\zeta(\alpha,x_{\min}) - \alpha\sum_{i=1}^{n} \log x_{i} \enspace ,
\end{align*}
which may be significantly more convenient than dealing with the derivative of the incomplete zeta function. This approach is what was used in both the present study, and in our preliminary study of this terrorism data~\citep{Clauset05}.

These equations assume that the range of the scaling behavior, i.e., the lower bound $x_{\min}$, is known. In real-world situations, this value is often estimated visually and a conservative estimate of such can be sufficient when the data span a half-dozen or so orders of magnitude. However, the data for many social or complex systems only span a few orders of magnitude at most, and an underpopulated tail would provide our tools with little statistical power. Thus, we use a numerical method for selecting the $x_{\min}$ that yields the best power-law model for the data. Specifically, for each $x_{\min}$ over some reasonable range, we first estimate the scaling parameter $\alpha_{\rm ML}$ over the data $x \geq x_{\min}$, and then compute the Kolmogorov-Smirnov (KS) goodness-of-fit statistic between the data being fit and a theoretical power-law distribution with parameters $\alpha_{\rm ML}$ and $x_{\rm min}$. We then select the $x_{\rm min}$ that yields the best such fit to our data. For simulated data with similar characteristics to the MIPT data, we find that this method correctly estimates both the lower bound on the scaling and the scaling exponent. Mathematically, we take
\begin{align*}
x_{\rm min} = \min_{y} \left[~\max_{x} \Big| F(x; \alpha_{\rm ML}, y) - \hat{F}(x; y) \Big|~\right] \enspace ,
\end{align*}
where $F(x; y, \alpha_{\rm ML})$ is the theoretical cumulative distribution function (cdf) for a power law with parameters $\alpha_{\rm ML}$ and $x_{\rm min}=y$, and $\hat{F}(x; y)$ is the empirical distribution function (edf) over the data points with value at least $y$. In cases where two values of $y$ yield roughly equally good fits to the data, we report the one with greater statistical significance.

Once these parameters have been estimated, we first calculate the standard error in $\alpha$ via bootstrap resampling. The errors reported in Tables 1 and 2, for instance, are derived in this manner. Finally, we calculate the statistical significance of this fit by a Monte Carlo simulation of $n$ data points drawn a large number of times (e.g., at least $1000$ draws) from $F(x;\alpha_{\rm ML}, x_{\rm min})$, where $\alpha_{\rm ML}$ and $x_{\rm min}$ have been estimated as above, under the one-sided KS test. Tabulating the results of the simulation yields an appropriate table of $p$-values for the fit, and by which the relative rank of the observed KS statistic can be interpreted in the standard way.

As mentioned in the text, there are many heavy-tailed distributions, e.g., the \mbox{q-exponential} $e_{q}^{-\alpha x}$, the stretched exponential $e^{-\alpha x^{\beta}}$, the log-normal, and even a different two-parameter power law $(c + x)^{-\alpha}$. For data that span only a few orders of magnitude, the behavior of these functions can be statistically indistinguishable, i.e., it can be hard to show that data generated from an alternative distribution would not yield just as good a fit to the power-law model. As such, we cannot rule out all Type II statistical errors for our power law models. On the other hand, we note that for the distributions described in Section~\ref{sec:fulldistribution}, the statistical power test versus a log-normal model indicates that the power law better represents the empirical data. In some sense, the particular kind of asymptotic scaling in the data is less significant than the robustness of the heavy tail under a variety of forms of analysis. Simply the fact that the patterns in the real-world severity data deviate so strongly from our expectations via traditional models of terrorism illustrates that there is much left to understand about this phenomenon, and our models need to be extended to account for the robust empirical patterns we observe in our study.

\end{appendix}

\singlespace
\fontsize{10}{10}
\selectfont

\bibliography{CYG_JCR}
\bibliographystyle{chicago}

\end{document}